\documentclass{article}

\usepackage[english]{babel}
\usepackage[hyphens,spaces]{url}

\usepackage[a4paper]{geometry}

\usepackage{amsmath}
\usepackage{graphicx}
\usepackage[colorlinks=true, allcolors=blue]{hyperref}

\title{Johnson's contribution to the Discussion of `Statistical aspects of the Covid-19 response' by Wood et al.}
\author{Oliver Johnson\thanks{School of Mathematics, 
	University of Bristol, Fry Building, Woodland Road, Bristol, 
	BS8 1UG, UK.
	Email: {\tt O.Johnson@bristol.ac.uk}
	} }

\begin{document}
\maketitle

As a data-driven `COVID centrist', conscious of the harms of both  disease and NPIs, I welcome this paper \cite{wood} and other attempts to calibrate the effectiveness of the UK response. However, I would like to raise some points :

\section{Section 3: Cases and ONS}
\subsection{Cases and prevalence}
\begin{figure}[ht!]
\centering
\includegraphics[width= 9cm]{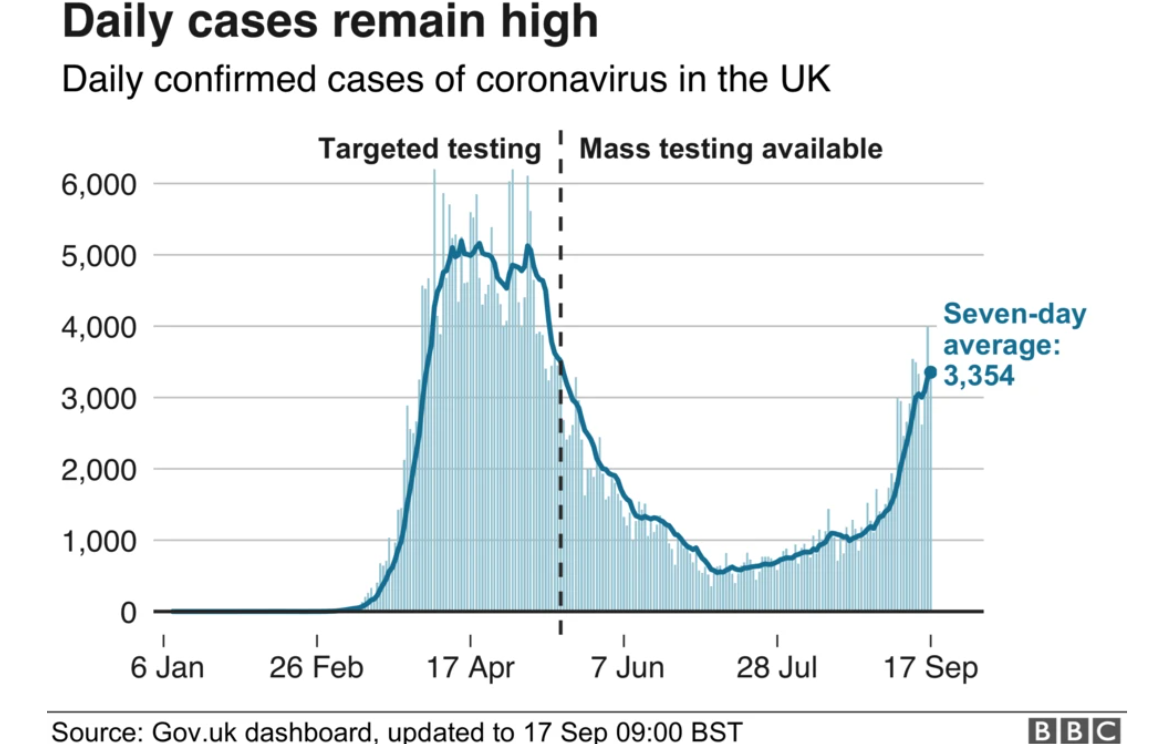}
\caption{Illustration of typical BBC graph of case data, taken from 18th September 2020 \cite{BBCstory}. \label{fig:BBCgraph}}
\end{figure}

While of course cases (positive tests) are not the same as prevalence (total infected people), I believe most media and commentators generally made that distinction clear. As an illustration, BBC graphs such as Figure \ref{fig:BBCgraph} typically (if somewhat arbitrarily) distinguished  `targeted testing' and `mass testing' regimes.

\begin{figure}[ht!]
\centering
\begin{minipage}{.45\textwidth}
\includegraphics[width= 6cm]{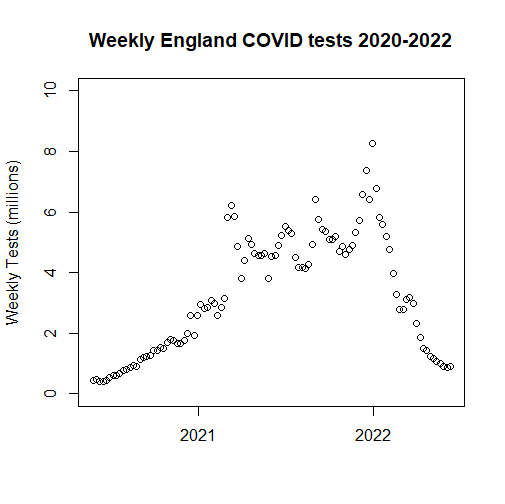}
\end{minipage}
\begin{minipage}{.45\textwidth}
\includegraphics[width= 6cm]{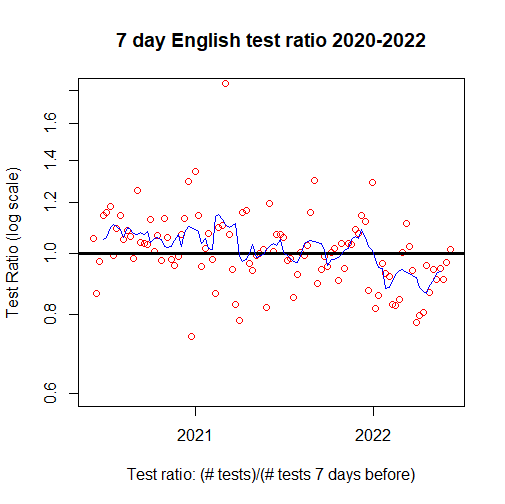}
\end{minipage}
\caption{Data on weekly tests in England (28th May 2020--9th June 2022) from \cite{testtrace}. a) Absolute weekly numbers (linear scale) 
b) Ratio of number of tests in a given week to number in previous week. Dots represent weekly change, line is smoothed average of three points. The single week with very high ratio (big increase in testing) corresponds to introduction of lateral flow tests in schools for March 2021 reopening.
 \label{fig:testratio}}
\end{figure}

The `curious argument' \cite[P.12]{wood} that short-term fluctuations in test numbers can be ignored to first approximation can be somewhat justified by plotting week-on-week ratios of test numbers \cite{testtrace} (Figure \ref{fig:testratio}). This shows that typical weekly changes in test numbers were on the order of 10\% or less, much smaller in magnitude than changes in infection numbers (see Figure \ref{fig:deaths} for death data), suggesting that estimates of trends in growth periods were broadly reliable.

\subsection{Reporting lag}
\begin{figure}[ht!]
\centering
\begin{minipage}{.5\textwidth}
\includegraphics[width = 6cm]{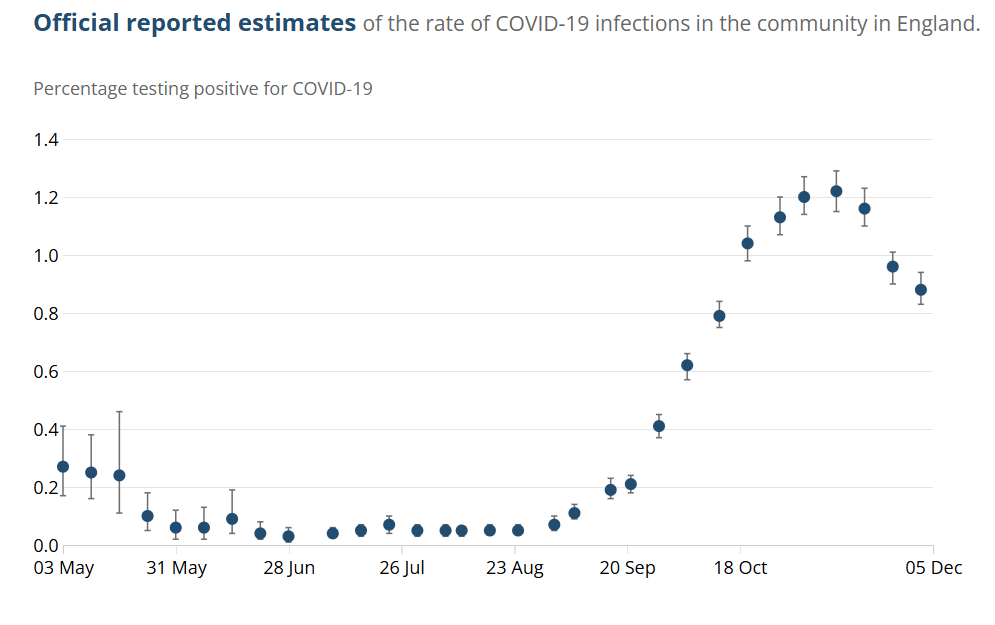}
\end{minipage}
\begin{minipage}{.4\textwidth}
\includegraphics[width = 4cm]{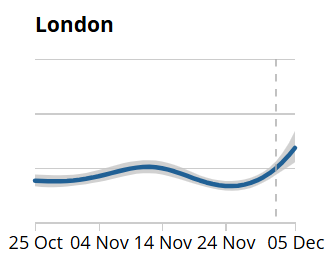}
\end{minipage}
\caption{Graphs of estimated COVID prevalence taken from ONS survey, 11th December 2020 a) England b) London \label{fig:ONS}}
\end{figure}

\begin{figure}[ht!]
\centering
\begin{minipage}{.45\textwidth}
\includegraphics[width = 5cm]{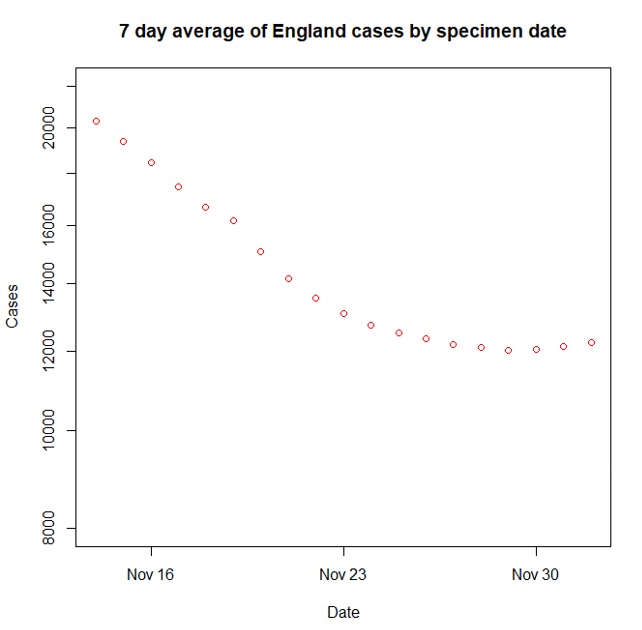}
\end{minipage}
\begin{minipage}{.45\textwidth}
\includegraphics[width = 5cm]{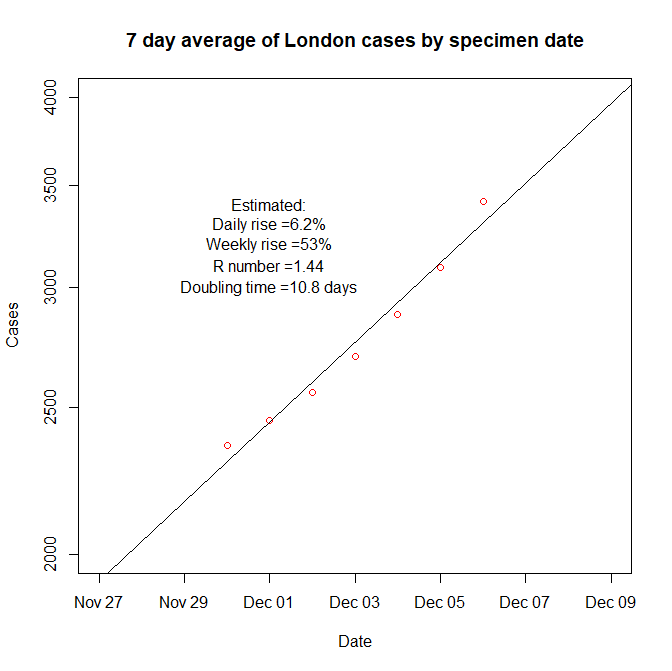}
\end{minipage}
\caption{Graphs tweeted by Oliver Johnson \makeatletter
@BristOliver \makeatother based on contemporaneous case data
a) 9th December 2020 (England) \cite{johnson091220} b) 14th December 2020 (London) \cite{johnson141220} \label{fig:johnson}}
\end{figure}

Further, while more precise, the ONS survey suffered from reporting lag. For example, on 11th December 2020 \cite{ONS111220} it reported an overall reduction in prevalence in England and only a modest rise in London (Figure \ref{fig:ONS}). These delayed estimates, referring to 29th November to 5th December, held for the next week and fed into the official consensus 11th--17th December R number estimates \cite{Rnumber} of 0.8-1.0 (England) and 0.9-1.1 (London). However contemporaneous analysis of cases, which are closer to the key incidence metric than to PCR prevalence,  gave more timely warning of the alpha wave (Figure \ref{fig:johnson} \cite{johnson091220,johnson141220}).

\newpage

\section{Section 5: Lockdowns}
While it's interesting to consider the possibility that action just short of a lockdown could cause infections to peak ($R$ to go under 1), a key statement \cite[P20]{wood} is `lockdowns having further suppressed infections, causing infection waves to subside more quickly than might have otherwise occurred'.

\begin{figure}[ht!]
\centering
\begin{minipage}{.45\textwidth}
\includegraphics[width = 6cm]{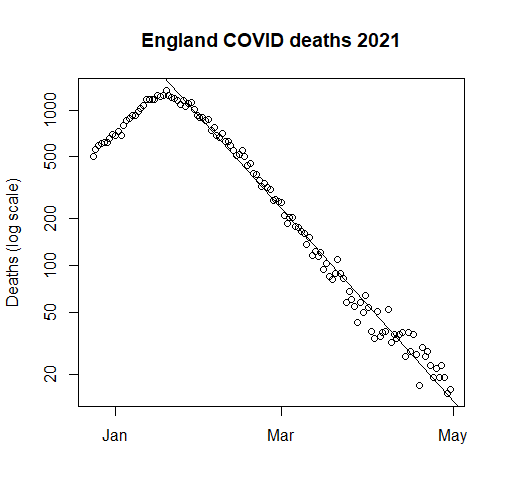}
\end{minipage}
\begin{minipage}{.45\textwidth}
\includegraphics[width = 6cm]{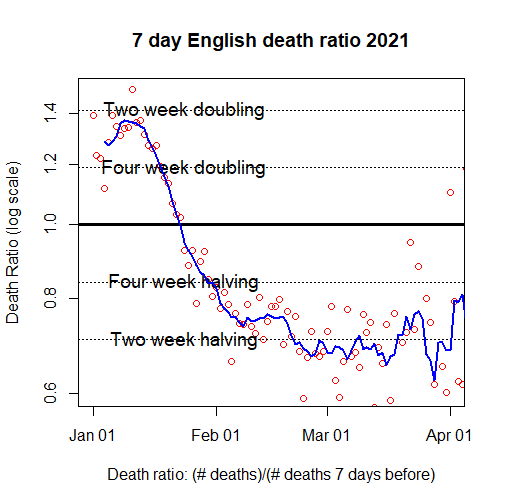}
\end{minipage}
\caption{England death data January--April 2021 \label{fig:deaths} by date of occurrence. a) Absolute numbers of deaths (log scale), line fitted to period of exponential decay using {\tt{lm()}} in {\tt R}. b) Daily death ratio (deaths on a given day divided by deaths 7 days before, to smooth out weekly effects).  Data from archived UKHSA COVID dashboard \cite{dashboard}.}
\end{figure}

To illustrate this, Figure \ref{fig:deaths} shows that under lockdown deaths in early 2021 were well modelled by a consistent exponential fall (straight line on log scale), with the fitted line representing a 28\% weekly decline in deaths and greater later variability due to random fluctuations.

\begin{figure}[ht!]
\centering
\includegraphics[width= 8cm]{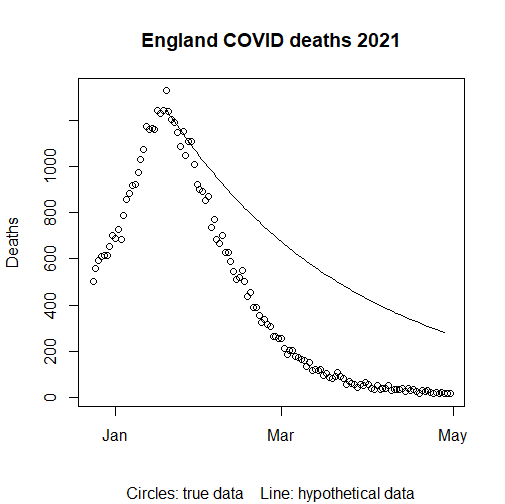}
\caption{True death data \cite{dashboard} plotted as circles (linear scale, estimated 28\% weekly decline), along with hypothetical trajectory plotted as a line (modelled 10\% weekly decline). \label{fig:hyp}}
\end{figure}

As a result, while deaths fell from a peak of 1,328 on 19th January, there were a total of 33,773 deaths in the next 100 days. Obviously, we cannot know what would have happened without lockdown, but presumably greater social mixing would have led to a slower rate of exponential decay. For illustration, even accepting the idea that $R<1$ might be possible without lockdown, Figure \ref{fig:hyp} shows a hypothetical trajectory of a slower (10\% weekly) decline resulting in 65,100 deaths over the same period.

Applying the NICE criteria of \cite[Section 1]{wood}, it would be reasonable to spend \pounds9bn to avert these extra $\sim$30,000 deaths, with presumably some significant additional monetary value from reduced hospitalizations (including greater available hospital capacity to treat other conditions). While this figure may be lower than the average \pounds3.7bn/month spent on the furlough scheme alone \cite{furlough}, it's not hugely different (and obviously different hypothetical rates of decline would lead to different figures).

Further, it illustrates a general principle that the benefits of lockdown are likely to decay exponentially over time, while costs accumulate linearly \cite[Figure 2]{wood}. Indeed, 32,052 of these 33,773 deaths occurred in the first 60 days. This crude analysis might perhaps suggest that the first two months of 2021 lockdown met the NICE criteria cited by \cite{wood}, but the benefit of later measures was less clear.

In general, such analysis might imply that the optimal strategy may involve short and intense periods of maximal suppression -- giving weight to the conclusion of \cite[P.20]{wood} that `the decision to continue suppression well into the summer of 2020 does not seem optimal'.  However, unlike \cite[Figure 1]{wood}, it feels that any analysis of deaths and consequences should take into account that very high numbers of COVID deaths were concentrated into relatively short periods of time.


\clearpage

\begin{thebibliography}{1}

\bibitem{BBCstory}
{BBC News}.
\newblock Coronavirus: 'widespread virus growth across the country', 18th
  September 2020.
\newblock \url{https://www.bbc.co.uk/news/health-54206705}.

\bibitem{furlough}
{House of Commons Library}.
\newblock Coronavirus job retention scheme: statistics, 23rd December 2021.
\newblock
  \url{https://commonslibrary.parliament.uk/research-briefings/cbp-9152/}.

\bibitem{johnson091220}
O.~Johnson.
\newblock {Tweet}, 2020.
\newblock 9th December:
  \url{https://twitter.com/BristOliver/status/1336704276878516226}.

\bibitem{johnson141220}
O.~Johnson.
\newblock {Tweet}, 2020.
\newblock 14th December:
  \url{https://twitter.com/BristOliver/status/1338450287040876545}.

\bibitem{ONS111220}
{Office for National Statistics (ONS)}.
\newblock {Coronavirus (COVID-19) Infection Survey, UK}, 11th December 2020.
\newblock
  \url{https://www.ons.gov.uk/peoplepopulationandcommunity/healthandsocialcare/conditionsanddiseases/bulletins/coronaviruscovid19infectionsurveypilot/11december2020}.

\bibitem{testtrace}
{UK Health Security Agency (UKHSA)}.
\newblock {Weekly statistics for NHS Test and Trace (England): 2 to 15 June
  2022}, 15th June 2022.
\newblock
  \url{https://www.gov.uk/government/publications/weekly-statistics-for-nhs-test-and-trace-england-2-to-15-june-2022}.

\bibitem{Rnumber}
{UK Health Security Agency (UKHSA)}.
\newblock The {R} value and growth rate, 2022.
\newblock \url{https://www.gov.uk/guidance/the-r-value-and-growth-rate}.

\bibitem{dashboard}
{UK Health Security Agency (UKHSA)}.
\newblock {COVID-19 Archive} data download, 2025.
\newblock
  \url{https://ukhsa-dashboard.data.gov.uk/covid-19-archive-data-download}.

\bibitem{wood}
S.~N. Wood, E.~C. Wit, P.~M. McKeigue, D.~Hu, B.~Flood, L.~Corcoran, and
  T.~Abou~Jawad.
\newblock Some statistical aspects of the {Covid-19} response.
\newblock {\em Journal of the Royal Statistical Society Series A: Statistics in
  Society (to appear)}, 2025.

\end{thebibliography}

\end{document}